# Longitudinal Josephson effect in systems with pairing of spatially separated electrons and holes

S.I. Shevchenko, O.M. Konstantynov

*B. Verkin Institute for Low Temperature Physics and Engineering of NAS of Ukraine,*

*47 Nauky Ave., Kharkiv, 61103, Ukraine*

E-mail: shevchenko@ilt.kharkov.ua

Longitudinal non-dissipative current states in bilayer electron-hole systems in the presence of potential barriers that divide the system into left and right sides was investigated. The consideration is performed both for the case of weak carrier coupling (high density) and strong coupling (low density). It is shown that in the high-density limit, the critical current is proportional to the product of the transparencies of the barriers in the electron and hole layers, whereas in the low-density limit, the current is inversely proportional to the sum of the heights of the potential barriers.

Keywords: Josephson effect, electron-hole currents, order parameter phase.

**1. Introduction**

About fifty years ago, in works [1-3] it was predicted that in bilayer systems, where one layer has electron conductivity and the other has hole conductivity, pairing of spatially separated carriers and transition of the resulting pairs to a superfluid state can occur. Since the motion of such pairs generates a non-dissipative electric current in each layer, directed in mutually opposite sides, the predicted phenomenon was named counterflow superconductivity.

In development of the ideas of studies [1-3], in work [4] it was proposed to place a two-layer system in a strong magnetic field perpendicular to the layers and to implement carrier pairing belonging to the zeroth Landau level. Soon after, it was realized [5-6] that in the presence of a strong magnetic field normal to the layers, electron-hole pairing is possible not only in *n-p* systems, but also in systems with the same type of conductivity. In this case, it is necessary that the total filling factor of the zeroth Landau level be equal to one. At this filling factor, the number of electrons in one layer coincides with the number of unoccupied states (holes) in the other layer. In the zero-level Landau approximation, there is an exact transformation [7] that converts an electron-electron system with Landau level filling factors $\nu_1$ and $\nu_2$ in layers 1 and 2, respectively, into an electron-hole system with layer filling factors $\nu_e = \nu_1$ and $\nu_h = 1 - \nu_2$. This means that in the case of a perpendicular magnetic field, when $\nu_1 + \nu_2 = 1$, pairing in bilayer electron systems is the same as in bilayer electron-hole systems. To date, a large number of experimental and theoretical studies have been carried out to investigate the behavior of bilayer systems with the same type of conductivity in each layer in a strong magnetic field perpendicular to the layers [8-14].

It should be noted that a strong magnetic field removes one significant constraint imposed by the carrier dispersion laws on the possibility of their pairing. This refers to high-density systems where the inequality $n_{pair} a_{pair}^2 \gg 1$ is satisfied, where $n_{pair}$ – electron-hole pair density, $a_{pair}$ – average pair size. For pairing to occur, the dispersion laws of the carriers $\varepsilon_i(\mathbf{p})$ must be almost congruent. In a strong magnetic field, the requirement for congruence is absent.

Without a magnetic field, the condition of congruence of the dispersion laws in high-density systems can be fulfilled, in particular, for systems formed by two layers of graphene separated by a thin (about 1*nm* thick) dielectric layer of hexagonal boron nitride (hBN) [15] and for systems consisting of a pair of parallel bilayer graphene sheets [16-20]. In the latter case the distance between the sheets is 0.37*nm*, which allows strong tunneling between the sheets. Due to this tunneling, in the density range of $4 \cdot 10^{12} cm^{-2} > n > 10^{11} cm^{-2}$, graphene bilayers behave like a zero-gap semiconductor with a quadratic dispersion law around the Fermi level $\varepsilon(\mathbf{k}) = \pm \hbar^2 k^2 / 2m^*$, where the effective mass is $m^* = (0.03 - 0.05) m_{0e}$, $m_{0e}$ – mass of a free electron.

A number of works investigate systems in which germanium layers are replaced by layers of transition metal dichalcogenides [21-25]. In particular, one layer is formed by $WSe_2$, and the other by $MoSe_2$, with an electron density that can reach $n = 10^{12} cm^{-2}$. It was reported [25] that in such a system, anomalous luminescence was observed, which is interpreted as a consequence of Bose condensation of interlayer excitons. The phenomenon was observed up to temperatures $100K$.

The congruence problem is absent in low-density electron systems ($n_{pair} a_{pair}^2 \ll 1$), but in such systems, with decreasing density, the transition temperature to the superfluid state decreases [26-31].

In conclusion of the introduction, we note the reviews [14, 32-35] and the works of the most recent years [36-38], which indicate that counterflow superconductivity in systems with pairing of spatially separated electrons and holes remains a relevant problem in condensed matter physics.

Systems in which pairing of spatially separated electrons and holes occurs, in fact represent a Josephson-like junction. The peculiarity of this junction lies in the fact that tunneling of carriers through a dielectric layer separating the conductive layers removes the degeneracy of the order parameter phase. A large number of studies are devoted to the issues that arise in this context [31, 39-48] (see also reviews [34], [35]). It should be noted that in the considered bilayer systems, two different Josephson effects can occur. The first one is realized due to the aforementioned interlayer tunneling. We will refer to this effect as the transverse Josephson effect and will not consider it in this work. The second effect (which we will call the longitudinal Josephson effect) occurs in a bilayer system where the right and left regions are weakly coupled (see. Fig. 1).

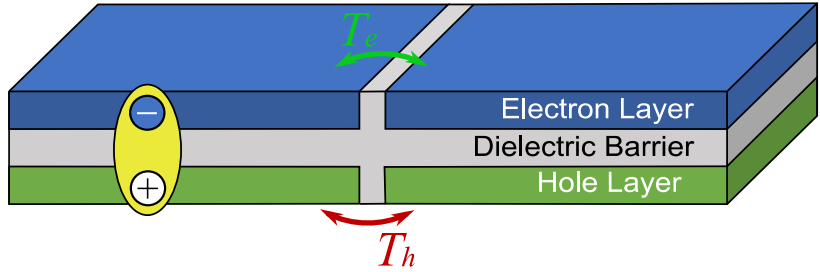

*Fig. 1.* Schematic representation of the bilayer system in which the right and left regions with electron-hole pairing are weakly coupled.

This work is devoted to the study of the longitudinal Josephson effect. The possibility of such an effect in high-density systems was pointed out in work [2], but many significant features of the effect remain unconsidered.

**2. High density limit**

Let us consider a bilayer electron-hole system consisting of flat parallel conducting layers with electron and hole conductivity types, respectively. We will assume that the layers are spaced sufficiently far apart so that transverse tunneling of carriers can be neglected. The Hamiltonian of such a system, in the second quantization representation, is written as:

$$H = \sum_{\mathbf{p},\sigma} \left[ \xi_{\mathbf{p}}^e a_{\mathbf{p},\sigma}^\dagger a_{\mathbf{p},\sigma} + \xi_{\mathbf{p}}^h b_{\mathbf{p},\sigma}^\dagger b_{\mathbf{p},\sigma} \right] + $$
$$+ \frac{1}{S} \sum_{\mathbf{p},\mathbf{p}_1,\mathbf{p}_2,\sigma} V_{eh}(p) a_{\mathbf{p}_1+\mathbf{p},\sigma}^\dagger b_{\mathbf{p}_2-\mathbf{p},-\sigma}^\dagger b_{\mathbf{p}_2,-\sigma} a_{\mathbf{p}_1,\sigma} \quad (1)$$

Here, $\xi_{\mathbf{p}}^e$ and $\xi_{\mathbf{p}}^h$ are the energies of the electron and hole, measured relative to their chemical potentials, $a_{\mathbf{p},\sigma}^{\dagger}$ and $b_{\mathbf{p},\sigma}^{\dagger}$ are the creation operators for electrons and holes in a state with momentum $\mathbf{p}$ and spin projection $\sigma = \uparrow, \downarrow$, $V_{eh}(p)$ is the Fourier component of the screened Coulomb interaction potential between the electron and the hole, and $S$ is the area of the system. The interaction between carriers within a layer, as in normal metals, leads to the renormalization of energies $\xi_{\mathbf{p}}^e$ and $\xi_{\mathbf{p}}^h$. The attraction between electrons and holes leads to the formation of bound electron-hole pairs. In the high-density limit, as in the Bardeen-Cooper-Schrieffer (BCS) theory, pairing is considered singlet, and the four-fermion Hamiltonian (1) is approximated by an effective Hamiltonian $H_{\textit{eff}}$, in which the creation and annihilation operators for electron-hole pairs are replaced by c-numbers [49]:

$$H_{\textit{eff}} = \sum_{\mathbf{p},\sigma} \left[ \xi_{\mathbf{p}}^e a_{\mathbf{p},\sigma}^{\dagger} a_{\mathbf{p},\sigma} + \xi_{\mathbf{p}}^h b_{\mathbf{p},\sigma}^{\dagger} b_{\mathbf{p},\sigma} \right] - \\ - \sum_{\mathbf{p}} \left[ \left( \Delta_{\mathbf{p}} b_{\mathbf{p},\uparrow}^{\dagger} a_{-\mathbf{p},\downarrow}^{\dagger} + h.c. \right) - \Delta_{\mathbf{p}} \left\langle a_{-\mathbf{p},\downarrow} b_{\mathbf{p},\uparrow} \right\rangle^* \right], \quad (2)$$

where the angle brackets denote averaging over the ground state of $H_{\textit{eff}}$, and $\Delta_{\mathbf{p}}$ is the order parameter, which is defined by the following expression:

$$\Delta_{\mathbf{p}} = -\frac{1}{S} \sum_{\mathbf{p}'} V_{eh}(\mathbf{p}-\mathbf{p}') \left\langle a_{-\mathbf{p}',\downarrow} b_{\mathbf{p}',\uparrow} \right\rangle. \quad (3)$$

To determine the energy levels of the system, the Hamiltonian $H_{\textit{eff}}$, as is known, should be reduced to a diagonal form, which is achieved by a linear transformation of the operators $a_{\mathbf{p},\sigma}$ and $b_{\mathbf{p},\sigma}$

$$\begin{cases} a_{\mathbf{p},\sigma} = u_{\mathbf{p}} \alpha_{\mathbf{p},\sigma} + v_{\mathbf{p}} \beta_{-\mathbf{p},-\sigma}^{\dagger} \\ b_{\mathbf{p},\sigma} = u_{\mathbf{p}} \beta_{\mathbf{p},\sigma} - v_{\mathbf{p}} \alpha_{-\mathbf{p},-\sigma}^{\dagger} \end{cases}. \quad (4)$$

The introduced quasiparticle operators $\alpha_{\mathbf{p},\sigma}$ and $\beta_{\mathbf{p},\sigma}$ satisfy the fermionic commutation relations. The functions $u_{\mathbf{p}}$ and $v_{\mathbf{p}}$ are determined from the condition that the off-diagonal terms in $H_{\textit{eff}}$ vanish. These functions have the following form:

$$u_{\mathbf{p}}^2 = \frac{1}{2}\left(1 + \frac{\xi_{\mathbf{p}}}{\varepsilon_{\mathbf{p}}}\right), \; v_{p}^2 = \frac{1}{2}\left(1 - \frac{\xi_{\mathbf{p}}}{\varepsilon_{\mathbf{p}}}\right), \quad (5)$$

where $\xi_{\mathbf{p}} = \left(\xi_{\mathbf{p}}^e + \xi_{\mathbf{p}}^h\right)/2$ and $\varepsilon_{\mathbf{p}} = \sqrt{\xi_{\mathbf{p}}^2 + \Delta_{\mathbf{p}}^2}$. As a result

$$H_{\textit{eff}} = \sum_{\mathbf{p},\sigma} \left[ \left(\varepsilon_{\mathbf{p}} + \eta_{\mathbf{p}}\right) \alpha_{\mathbf{p},\sigma}^{\dagger} \alpha_{\mathbf{p},\sigma} + \left(\varepsilon_{\mathbf{p}} - \eta_{\mathbf{p}}\right) \beta_{\mathbf{p},\sigma}^{\dagger} \beta_{\mathbf{p},\sigma} \right], \quad (6)$$

where $\eta_{\mathbf{p}} = \left(\xi_{\mathbf{p}}^e - \xi_{\mathbf{p}}^h\right)/2$. In expression (6), the c-number part of the ground state energy is omitted. The appearance of the energy $\eta_{\mathbf{p}}$ in (6) is due to the difference in the dispersion laws of electrons in the valence band $\xi_{\mathbf{p}}^h$ and in the conduction band $\xi_{\mathbf{p}}^e$. Substituting (4) into (3) and assuming that the quasiparticles are in a state of thermodynamic equilibrium at temperature $T$, that is

$$\left\langle \alpha_{\mathbf{p},\sigma}^{\dagger} \alpha_{\mathbf{p},\sigma} \right\rangle = \left[ \exp\left(\frac{\varepsilon_{\mathbf{p}} + \eta_{\mathbf{p}}}{T}\right) + 1 \right]^{-1} \equiv f_{\mathbf{p}}^e, \quad (7)$$

$$\left\langle \beta^{\dagger}_{\mathbf{p},\sigma} \beta_{\mathbf{p},\sigma} \right\rangle = \left[ \exp\left( \frac{\varepsilon_{\mathbf{p}} - \eta_{\mathbf{p}}}{T} \right) + 1 \right]^{-1} \equiv f^{h}_{\mathbf{p}}, \tag{8}$$

we obtain the self-consistency equation for the order parameter

$$\Delta_{\mathbf{p}} = -\frac{1}{S} \sum_{\mathbf{p}'} \frac{\Delta_{\mathbf{p}'} V_{eh}(\mathbf{p}-\mathbf{p}')}{2\varepsilon_{\mathbf{p}'}} \left( 1 - f^{e}_{\mathbf{p}'} - f^{h}_{\mathbf{p}'} \right). \tag{9}$$

As mentioned in the introduction, in the absence of an external magnetic field, the congruence of the Fermi surfaces of electrons and holes is critically important for the realization of their pairing. The incongruence will lead to a significant reduction in the range of allowable interlayer distances $d$ and carrier densities $n$, at which equation (9) has a nontrivial solution. This issue for the case of an excitonic dielectric was discussed in more detail in [50]. The case of the considered bilayer system is analogous to the situation with the formation of an excitonic phase in a semimetal [50], where the densities of electrons and holes coincide, but the feature of the bilayer system is the presence of an additional parameter $d$.

Let us demonstrate this by considering the example of weak incongruence of the Fermi surfaces of electrons and holes for the case of zero temperature, where the factor in parentheses (9) becomes equal to one. Consider the case of the isotropic dispersion law of electrons

$$\xi^{e}_{\mathbf{p}} = \frac{p^2 - p_0^2}{2m_e}, \tag{10}$$

whereas the dispersion law of holes has a weak anisotropy

$$\xi^{h}_{\mathbf{p}} = \frac{p_x^2 + \gamma^2 p_y^2 - \gamma p_0^2}{2m_h}, \tag{11}$$

expressed by the presence of the factor $\gamma < 1$ such that $1 - \gamma \ll 1$. Here $m_e$ and $m_h$ are the effective masses of the electron and hole, respectively, $p_0$ is the Fermi momentum, which determines the density of electrons and holes $n_e = n_h = p_0^2 / 2\pi\hbar^2 \equiv n$, and the $xy$-plane is parallel to the conducting layers. For the case of zero temperature, the linearized equation (9) takes the form

$$\Delta_{\mathbf{p}} = -\frac{1}{(2\pi\hbar)^2} \int \frac{\Delta_{\mathbf{p}'} V_{eh}(\mathbf{p}-\mathbf{p}')}{\left| \xi^{e}_{\mathbf{p}'} + \xi^{h}_{\mathbf{p}'} \right|} d\mathbf{p}'. \tag{12}$$

Substituting (10) and (11) into (12), taking into account the small parameter $1-\gamma$ and neglecting the dependence of the order parameter $\Delta_{\mathbf{p}}$ on the angular variable, we obtain the equation for the critical line that separates the region of the excitonic phase from the region where $\Delta_{\mathbf{p}} \equiv 0$. This equation has the form

$$1 = \ln\left[ 1/\left(1-\gamma^2\right) \right] \frac{m}{\pi\hbar^2} \int_{0}^{2\pi} \frac{\left| V_{eh}\left(2p_0 \sin(\theta/2)\right) \right|}{2\pi} d\theta, \tag{13}$$

where $m = m_e m_h / M$ is the reduced mass of the electron and hole, $M = m_e + m_h$. After substituting into (13) the expression for the screened Coulomb interaction potential

$$V_{eh}(p < 2p_0) = -\frac{\pi\hbar^2 e}{M} \frac{\exp(-pd/\hbar)}{1 + \dfrac{pa_0}{2\hbar}\dfrac{m}{M} + \dfrac{2\hbar}{pa_0}\left(1 - e^{-\frac{2pd}{\hbar}}\right)}, \tag{14}$$

obtained using the random phase approximation, where $a_0 = \hbar^2 \varepsilon / me^2$ is the effective Bohr radius of an electron-hole pair, for $1-\gamma^2 = 10^{-4}$ and $m_e = m_h$, we find the graph of the dependence of the density $n$ on the interlayer distance $d$ (see Fig. 2), that is, the aforementioned critical line.

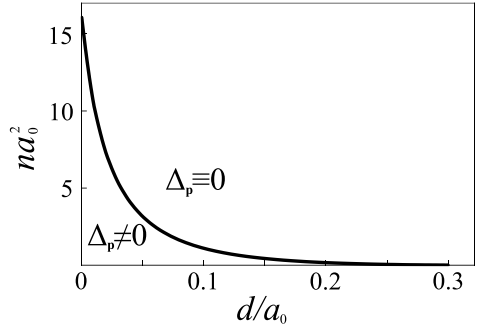

*Fig. 2.* The critical curve, separating the region of the excitonic phase $\Delta_\mathbf{p} \neq 0$ from the region where $\Delta_\mathbf{p} \equiv 0$ for the case of an isotropic electron dispersion law and an anisotropic hole dispersion law with an anisotropy factor $1-\gamma^2 = 10^{-4}$ for $m_e = m_h$ and $T = 0$.

Thus, even weak incongruence of the Fermi surfaces of electrons and holes can be critically important for the formation of the superfluid excitonic phase. Further, we will limit ourselves to the case of isotropic dispersion laws.

Let us proceed to the study of the longitudinal Josephson effect in the considered bilayer system. In the work [2], results obtained using a method called the *t*-representation method (see also [51]), are presented without derivation; some details of this method are provided in the Appendix. Here, we will use the well-known method of the tunneling Hamiltonian. If $H_1$ and $H_2$ are the Hamiltonians of the left and right half-spaces, corresponding to the isolated bilayer systems, and the term $H_T$ describes the tunneling transitions of electrons and holes across the barrier, then the Hamiltonian of the system can be written as follows:

$$H = H_1 + H_2 + H_T. \tag{15}$$

Here, $H_1$ and $H_2$ are given by expressions of the form (6), and $H_T$ is equal to

$$H_T = \sum_{\mathbf{p},\mathbf{q},\sigma} \left( T^e_{\mathbf{p},\mathbf{q}} a^\dagger_{1\mathbf{p},\sigma} a_{2\mathbf{q},\sigma} + T^h_{\mathbf{p},\mathbf{q}} b^\dagger_{1\mathbf{p},\sigma} b_{2\mathbf{q},\sigma} + h.c. \right). \tag{16}$$

Here, $a^\dagger_{1\mathbf{p},\sigma}$ and $b^\dagger_{1\mathbf{p},\sigma}$ are the creation operators for left electrons and holes in the state with momentum $\mathbf{p}$ and spin projection $\sigma = \uparrow, \downarrow$; $a^\dagger_{2\mathbf{q},\sigma}$ и $b^\dagger_{2\mathbf{q},\sigma}$ are the corresponding operators for right electrons and holes; $T^e_{\mathbf{p},\mathbf{q}}$ and $T^h_{\mathbf{p},\mathbf{q}}$ are matrix elements defining the tunneling probability for electrons and holes, respectively.

In the tunneling Hamiltonian method, the current, for example, in the electron layer $I_e$, is determined by the rate of change of the number of electrons in one of the half-spaces:

$$I_e = e\left\langle \frac{d}{dt} \sum_{\mathbf{p},\sigma} a^\dagger_{1\mathbf{p},\sigma} a_{1\mathbf{p},\sigma} \right\rangle = \frac{2e}{\hbar} \operatorname{Im} \sum_{\mathbf{p},\mathbf{q},\sigma} T^e_{\mathbf{p},\mathbf{q}} \left\langle \tilde{a}^\dagger_{1\mathbf{p},\sigma} \tilde{a}_{2\mathbf{q},\sigma} \right\rangle \exp\left( \frac{i(\varphi_1 - \varphi_2)}{2} \right), \tag{17}$$

where $e > 0$ is the elementary electric charge. Here, we have extracted the order parameter phases $\varphi_1$ and $\varphi_2$, corresponding to the left and right parts of the bilayer system.

$$a_{1,2\mathbf{p},\sigma} = \tilde{a}_{1,2\mathbf{p},\sigma} \exp\left( -\frac{i\varphi_{1,2}}{2} \right), \quad b_{1,2\mathbf{p},\sigma} = \tilde{b}_{1,2\mathbf{p},\sigma} \exp\left( -\frac{i\varphi_{1,2}}{2} \right). \tag{18}$$

In order to find the off-diagonal correlation function $\left\langle \tilde{a}^\dagger_{1\mathbf{p},\sigma} \tilde{a}_{2\mathbf{q},\sigma} \right\rangle$ appearing in (17), we perform the $u$–$v$-Bogoliubov transformation (4) on the operators $\tilde{a}^\dagger_{1\mathbf{p},\sigma}$ and $\tilde{a}_{2\mathbf{q},\sigma}$. As a result,

$$\left\langle \tilde{a}^\dagger_{1\mathbf{p},\sigma} \tilde{a}_{2\mathbf{q},\sigma} \right\rangle = u_p u_q \left\langle \alpha^\dagger_{1\mathbf{p},\sigma} \alpha_{2\mathbf{q},\sigma} \right\rangle + v_p v_q \left\langle \beta_{1-\mathbf{p},-\sigma} \beta^\dagger_{2-\mathbf{q},-\sigma} \right\rangle + \\ + u_p v_q \left\langle \alpha^\dagger_{1\mathbf{p},\sigma} \beta^\dagger_{2-\mathbf{q},-\sigma} \right\rangle + v_p u_q \left\langle \beta_{1-\mathbf{p},-\sigma} \alpha_{2\mathbf{q},\sigma} \right\rangle \tag{19}$$

Here, as before, indices 1 and 2 refer to the left and right half-spaces, respectively. Using (4) and (6), one can construct the Heisenberg equations for the quasiparticle correlators entering (19). By neglecting off-diagonal correlators of the type $\langle \alpha^\dagger_{1\mathbf{p},\sigma} \tilde{a}_{1\mathbf{p}',\sigma} \rangle$, $\mathbf{p} \neq \mathbf{p}'$ (their inclusion leads to terms of higher order in transparency) and taking into account (7) and (8), we can easily obtain the solutions of these equations. For the case of zero voltage on the structure, they have the form

$$\langle \alpha^\dagger_{1\mathbf{p},\sigma} \alpha_{2\mathbf{q},\sigma} \rangle = (f^e_q - f^e_p) \left[ \frac{T^{h*}_{\mathbf{p},\mathbf{q}} v_p v_q \exp(i\varphi/2) - T^{e*}_{\mathbf{p},\mathbf{q}} u_p u_q \exp(-i\varphi/2)}{\varepsilon^e_p - \varepsilon^e_q + i\delta} \right], \quad (20)$$

$$\langle \beta_{1-\mathbf{p},-\sigma} \beta^\dagger_{2-\mathbf{q},-\sigma} \rangle = (f^h_q - f^h_p) \left[ \frac{T^{h*}_{\mathbf{p},\mathbf{q}} u_p u_q \exp(i\varphi/2) - T^{e*}_{\mathbf{p},\mathbf{q}} v_p v_q \exp(-i\varphi/2)}{\varepsilon^h_p - \varepsilon^h_q - i\delta} \right], \quad (21)$$

$$\langle \alpha^\dagger_{1\mathbf{p},\sigma} \beta^\dagger_{2-\mathbf{q},-\sigma} \rangle = (f^e_p + f^h_q - 1) \left[ \frac{T^{h*}_{\mathbf{p},\mathbf{q}} v_p u_q \exp(i\varphi/2) + T^{e*}_{\mathbf{p},\mathbf{q}} u_p v_q \exp(-i\varphi/2)}{\varepsilon^e_p + \varepsilon^h_q + i\delta} \right], \quad (22)$$

$$\langle \beta_{1-\mathbf{p},-\sigma} \alpha_{2\mathbf{q},\sigma} \rangle = (f^h_p + f^e_q - 1) \left[ \frac{T^{h*}_{\mathbf{p},\mathbf{q}} u_p v_q \exp(i\varphi/2) + T^{e*}_{\mathbf{p},\mathbf{q}} v_p u_q \exp(-i\varphi/2)}{\varepsilon^h_p + \varepsilon^e_q - i\delta} \right]. \quad (23)$$

Here, the notations $\varphi = \varphi_1 - \varphi_2$, $\varepsilon^e_p = \varepsilon_p + \eta_p$ and $\varepsilon^h_p = \varepsilon_p - \eta_p$ are introduced, and the value $\delta \to +0$. Substituting (20)–(23) into (19) and then into (17), considering (5), gives (compare with [51])

$$I_e = \frac{e}{\hbar} \operatorname{Im} \left\{ \exp(i\varphi) \sum_{\mathbf{p},\mathbf{q}} \frac{\Delta_p \Delta_q}{\varepsilon_p \varepsilon_q} T^e_{\mathbf{p},\mathbf{q}} T^{h\,*}_{\mathbf{p},\mathbf{q}} \times \right.$$
$$\left. \times \left[ \frac{f^e_q - f^e_p}{\varepsilon^e_p - \varepsilon^e_q + i\delta} + \frac{f^h_q - f^h_p}{\varepsilon^h_p - \varepsilon^h_q - i\delta} + \frac{f^e_p + f^h_q - 1}{\varepsilon^e_p + \varepsilon^h_q + i\delta} + \frac{f^h_p + f^e_q - 1}{\varepsilon^h_p + \varepsilon^e_q - i\delta} \right] \right\}. \quad (24)$$

Taking into account the relation $T^{e,h}_{\mathbf{p},\mathbf{q}} = T^{e,h\,*}_{-\mathbf{p},-\mathbf{q}}$, it can be shown that the expression inside the curly brackets, except for the phase factor $\exp(i\varphi)$, is real. As a result, by also using the representation of the Fermi-Dirac distribution function $f^{e,h}_p = T \sum_\omega (i\hbar\omega - \varepsilon^{e,h}_p)$ (where $\hbar\omega = \pi T(2n+1)$, $n$ – is an integer), after several simplifications, we obtain the final expression for the Josephson current

$$I_e = -\frac{2e \sin\varphi}{\hbar} \sum_{\mathbf{p},\mathbf{q},\omega} T \frac{T^e_{\mathbf{p},\mathbf{q}} T^{h\,*}_{\mathbf{p},\mathbf{q}} \Delta_p \Delta_q}{\left( \varepsilon^2_p + (\hbar\omega + i\eta_p)^2 \right)\left( \varepsilon^2_q + (\hbar\omega + i\eta_q)^2 \right)}. \quad (25)$$

Since the obtained current is proportional to the product of the tunneling matrix elements $T^e_{\mathbf{p},\mathbf{q}}$ and $T^h_{\mathbf{p},\mathbf{q}}$ through the barrier in the electron and hole layers, respectively, then for the longitudinal Josephson effect to occur in a high-density electron-hole bilayer system, the presence of weak coupling in both layers is required. It is also noteworthy that, compared to a conventional superconductor, the presence of the energy term $\eta_p$ in the denominator (25) is new; this term is associated with the asymmetry in the dispersion laws of electrons and holes by masses. For equal carrier masses and identical transparencies $T^e_{\mathbf{p},\mathbf{q}}$ and $T^h_{\mathbf{p},\mathbf{q}}$, the expression for the current $I_e$ coincides with that in conventional superconductors.

### 3. Low density limit

Let us now turn to the consideration of a dilute electron-hole gas in a bilayer system, which is described by a Hamiltonian of the form

$$H = \sum_{\substack{\alpha=e,h \\ \sigma=\uparrow,\downarrow}} \int d\mathbf{r}\, \psi_\sigma^{(\alpha)\dagger}(\mathbf{r}) \left[ -\frac{\hbar^2}{2m_\alpha}\nabla^2 + U_\alpha(\mathbf{r}) \right] \psi_\sigma^{(\alpha)}(\mathbf{r}) +$$
$$+ \frac{1}{2} \sum_{\substack{\alpha,\beta=e,h \\ \sigma,\sigma'=\uparrow,\downarrow}} \int d\mathbf{r}d\mathbf{r}'\, \psi_\sigma^{(\alpha)\dagger}(\mathbf{r})\psi_{\sigma'}^{(\beta)\dagger}(\mathbf{r}')V_{\alpha\beta}(|\mathbf{r}-\mathbf{r}'|)\psi_{\sigma'}^{(\beta)}(\mathbf{r}')\psi_\sigma^{(\alpha)}(\mathbf{r})$$
(26)

Here, $\psi_e^{(\alpha)\dagger}$ and $\psi_h^{(\alpha)\dagger}$ are the field creation operators, $m_e$ and $m_h$ are the effective masses of the electron and hole, respectively, $U_\alpha(\mathbf{r})$ is the external potential of the layer, $V_{\alpha\beta}$ is the energy of the Coulomb interaction between carriers, $\sigma = \uparrow, \downarrow$ is the spin index, $\mathbf{r}$ is the 2D radius vector. Denoting the thickness of the insulating layer between the electron and hole layers as $d$ and dielectric permittivity of the system as $\varepsilon$, we have $V_{ee}(r) = V_{hh}(r) = e^2/\varepsilon r$ and $V_{eh}(r) = V_{he}(r) = -e^2/\varepsilon\sqrt{r^2 + d^2}$.

The ground state wave function of the system $|\Phi_0\rangle$, following Keldysh, will be sought in the form of a generalized coherent state [52]:

$$|\Phi_0\rangle = \exp\{\sum_{\sigma,\sigma'=\uparrow,\downarrow} \int d\mathbf{r}_1 d\mathbf{r}_2 \Phi_{\sigma\sigma'}(\mathbf{r}_1,\mathbf{r}_2) e^{-i\mu_{\sigma\sigma'} t/\hbar} \psi_\sigma^{(e)\dagger}(\mathbf{r}_1)\psi_{\sigma'}^{(h)\dagger}(\mathbf{r}_2) - h.c.\}|0\rangle, \quad (27)$$

where $\mu_{\sigma\sigma'}$ is the chemical potential of the component of the electron-hole gas, described by the function $\Phi_{\sigma\sigma'}(\mathbf{r}_1,\mathbf{r}_2)$. In the following, we will consider singlet pairing of carriers, corresponding to a spin-diagonal matrix $\Phi_{\sigma\sigma'}(\mathbf{r}_1,\mathbf{r}_2)$. Furthermore, we will assume that the system has a phase separation such that in the matrix $\Phi_{\sigma\sigma'}(\mathbf{r}_1,\mathbf{r}_2)$ only the function $\Phi_{\uparrow\uparrow}(\mathbf{r}_1,\mathbf{r}_2) \equiv \Phi(\mathbf{r}_1,\mathbf{r}_2)$ is non-zero. As shown in the work [28], such a situation occurs when $d > 0.2a_0$, where $a_0$ is the effective Bohr radius of the pair.

The equation for the function $\Phi(\mathbf{r}_1,\mathbf{r}_2)$ is obtained from the condition of minimizing the free energy functional $F \equiv E - \mu N$, where $E$ and $N$ are the energy of the system and the number of electron-hole pairs, respectively, and $\mu \equiv \mu_{\uparrow\uparrow}$. This functional is expanded in even powers of $\Phi$. In the low-density limit, the equation for $\Phi$ is [28]

$$\left[ -\frac{\hbar^2}{2m_e}\frac{\partial^2}{\partial \mathbf{r}_1^2} - \frac{\hbar^2}{2m_h}\frac{\partial^2}{\partial \mathbf{r}_2^2} + U_e(\mathbf{r}_1) + U_h(\mathbf{r}_2) + V_{eh}(|\mathbf{r}_1 - \mathbf{r}_2|) - \mu \right]\Phi(\mathbf{r}_1,\mathbf{r}_2) +$$
$$+ \int V_d(\mathbf{r}_1,\mathbf{r}_2,\mathbf{r}_3,\mathbf{r}_4)\Phi(\mathbf{r}_1,\mathbf{r}_2)|\Phi(\mathbf{r}_3,\mathbf{r}_4)|^2 d\mathbf{r}_3 d\mathbf{r}_4 -$$
$$- \int V_{ex}(\mathbf{r}_1,\mathbf{r}_2,\mathbf{r}_3,\mathbf{r}_4)\Phi(\mathbf{r}_1,\mathbf{r}_4)\Phi^*(\mathbf{r}_3,\mathbf{r}_4)\Phi(\mathbf{r}_3,\mathbf{r}_2) d\mathbf{r}_3 d\mathbf{r}_4 = 0$$
(28)

Here, $V_d$ and $V_{ex}$ are the kernels of the direct and exchange Coulomb interactions, respectively.

$$V_d = V_{ee}(r_{13}) + V_{hh}(r_{24}) + V_{eh}(r_{14}) + V_{eh}(r_{23}), \quad (29)$$

$$V_{ex} = V_{ee}(r_{13}) + V_{hh}(r_{24}) + \frac{1}{2}[V_{eh}(r_{12}) + V_{eh}(r_{34}) + V_{eh}(r_{14}) + V_{eh}(r_{23})], \quad (30)$$

where $r_{ij} \equiv |\mathbf{r}_i - \mathbf{r}_j|$. The solution of equation (28) is sought in the form $\Phi(\mathbf{r}_1,\mathbf{r}_2) = \Psi(\mathbf{R}_{12})\phi_0(r_{12})$, where $\mathbf{R}_{12}$ is the coordinates of the center of mass of the electron-hole pair, and $\phi_0(r_{12})$ is the ground-state wave function of a single electron-hole pair, normalized to unity. As a result, equation (28) can be rewritten as follows:

$$\left[-\frac{\hbar^2}{2M}\frac{\partial^2}{\partial \mathbf{R}_{12}^2}-\tilde{\mu}+\int\left(U_e\left(\mathbf{R}_{12}+\frac{m_h}{M}\mathbf{r}\right)+U_h\left(\mathbf{R}_{12}-\frac{m_e}{M}\mathbf{r}\right)\right)\phi_0^2(r)d\mathbf{r}\right]\Psi(\mathbf{R}_{12})+$$
$$+\int V_d\,\phi_0^2(r_{12})\phi_0^2(r_{34})\Psi(\mathbf{R}_{12})|\Psi(\mathbf{R}_{34})|^2 d\mathbf{r}_3 d\mathbf{r}_4 d\mathbf{r}_{12}-$$
$$-\int V_{ex}\phi_0(r_{12})\phi_0(r_{23})\phi_0(r_{34})\phi_0(r_{14})\Psi(\mathbf{R}_{32})\Psi^*(\mathbf{R}_{34})\Psi(\mathbf{R}_{14})d\mathbf{r}_3 d\mathbf{r}_4 d\mathbf{r}_{12}=0 \quad (31)$$

Here, $M = m_e + m_h$, and $\tilde{\mu}$ is the chemical potential, measured from the ground-state energy of an isolated pair. Since the characteristic lengths over which the potentials $U_\alpha(\mathbf{r})$ change are much greater than the lengths over which the function $\phi_0(r)$ changes, in the integrand within the square brackets of (31) we can expand $U_\alpha$ near the point $\mathbf{r}=0$. For the same reason, in the integrals of direct and exchange interactions (31) we expand with respect to $\mathbf{r}_i$ the slowly varying functions $\Psi$ near $\mathbf{R}_{12}$. In the zeroth order of these expansions, equation (31) reduces to the Gross-Pitaevskii equation

$$\left[-\frac{\hbar^2}{2M}\frac{\partial^2}{\partial \mathbf{R}^2}-\tilde{\mu}+U_e(\mathbf{R})+U_h(\mathbf{R})\right]\Psi(\mathbf{R})+\gamma\,\Psi(\mathbf{R})|\Psi(\mathbf{R})|^2=0. \quad (32)$$

Here, $\gamma = \gamma_d + \gamma_{ex}$ is the interaction constant, which includes contributions from the direct $\gamma_d$ and exchange $\gamma_{ex}$ interactions

$$\gamma_d = \frac{4\pi e^2 d}{\varepsilon}, \quad (33)$$

$$\gamma_{ex} = -\frac{4\pi e^2}{\varepsilon}\int\frac{d\mathbf{p}d\mathbf{q}}{(2\pi)^4}\frac{|\phi_\mathbf{q}|^2}{p}\left[|\phi_{\mathbf{q}+\mathbf{p}}|^2-\frac{e^{-pd}}{2}\left(\phi^*_{\mathbf{q}+\mathbf{p}}\phi_\mathbf{q}+\phi^*_\mathbf{q}\phi_{\mathbf{q}+\mathbf{p}}\right)\right], \quad (34)$$

where $\phi_\mathbf{q}$ is the Fourier component of the wave function $\phi_0(r)$.

Thus, the problem of the longitudinal Josephson effect for the superfluid state of the system of spatially separated electrons and holes in the low-density limit reduces to the problem of tunneling of a Bose gas through a barrier with potential $U(\mathbf{R}) = U_e(\mathbf{R}) + U_h(\mathbf{R})$. We will assume that the potential $U$, and therefore the wave function $\Psi$, depend only on the longitudinal coordinate $x$. We will seek a solution for the wave function $\Psi(x)$ in the form $\Psi(x) = \sqrt{n}f(x)\exp(i\theta(x))$ ($n$ – is the equilibrium density of the system far from the barrier), which corresponds to the stationary Josephson current density

$$j = f^2\frac{d\theta}{dx} = const. \quad (35)$$

Here and below, the coordinate $x$ is measured in units of the coherence length $\xi = \hbar/\sqrt{2Mn\gamma}$, and the current is measured in units of $j_0 = \hbar n/M\xi$. According to (32), the function $f(x)$ satisfies the equation

$$f'' - \frac{j^2}{f^3} + \left(\frac{\tilde{\mu}}{\gamma n} - f^2\right)f = \frac{U(x)}{\gamma n\xi}f. \quad (36)$$

The chemical potential $\tilde{\mu}$ in this equation is determined from the boundary condition $f(\infty)=1$ and is equal to

$$\tilde{\mu} = \gamma n(1+j^2). \quad (37)$$

To solve equation (36), we approximate the true barrier potential with a delta-function potential, such that

$$U(x) = (U_{0e} + U_{0h})\delta(x). \tag{38}$$

The solution to equation (36) with the barrier (38) for the case of small currents $j \ll 1$, such that $\tilde{\mu} \approx \gamma n$, was considered in [53]. As will be shown below, this situation occurs when the left and right regions with electron-hole pairing are weakly coupled, so that the condition $(U_{0e} + U_{0h})/\gamma n\xi \equiv \lambda \gg 1$ is satisfied. In this approximation, the solution of equation (36) takes the form

$$f = \sqrt{2j^2 + \tanh^2\left(\frac{|x| + x_0}{\sqrt{2}}\right)}, \tag{39}$$

where the quantity $x_0$ is determined by the value of the function $f$ at zero, which can be obtained from the boundary condition to equation (36) at the point $x = 0$

$$2\frac{df_0}{dx} = \lambda f_0, \tag{40}$$

where $f_0 = f(+0)$. To solve this equation, it is more convenient to make the substitution $z = f^2$. As a result of substituting (39) into (40), we get the equation for $z_0 = z(0)$

$$\lambda^2 z_0^2 - 2z_0 + 4j^2 = 0. \tag{41}$$

The current $j$ is related to the phase of the order parameter $\theta$ by the expression (35). Substituting the function $f(x)$ obtained above into this expression and assuming $\theta(0) = 0$, we get:

$$\theta(x) = jx + \text{sgn}(x)\left[\arctan\left(\sqrt{\frac{z(x) - 2j^2}{2j^2}}\right) - \arctan\left(\sqrt{\frac{z_0 - 2j^2}{2j^2}}\right)\right]. \tag{42}$$

Taking into consideration that in the absence of a barrier $\theta = jx$, we introduce a function $\varphi(x) \equiv \theta(x) - jx$, which will represent a part of the phase associated with tunneling [54]. The dependencies of $\varphi(x)$ and $f(x)$ are shown in Fig. 3(a) and Fig. 3(b), respectively.

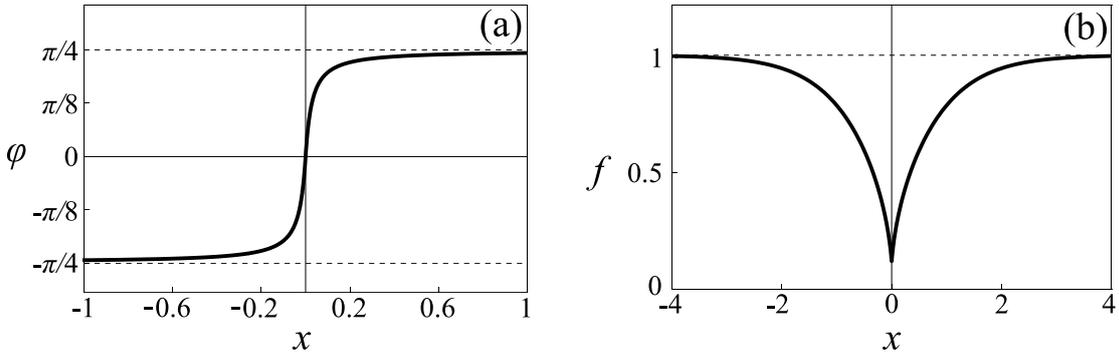

*Fig. 3.* The coordinate dependencies of the phase $\varphi$, associated with tunneling (a), and the function $f$, defining the system's density (b), for the case of low transparency ($\lambda = 50$) and critical current $j = 1/2\lambda$ (see (44)).

Define $\Delta\varphi$ as the phase difference between the right and left sides $\Delta\varphi = \varphi(+\infty) - \varphi(-\infty)$. Then, from (42), we obtain:

$$\Delta\varphi = \pi - 2\arctan\sqrt{\frac{z_0 - 2j^2}{2j^2}}. \tag{43}$$

From (41) and (43) follows the relationship between the tunneling current $j$ and the phase difference $\Delta\varphi$

$$j = \frac{1}{2\lambda} \sin \Delta\varphi. \qquad (44)$$

As can be seen from this relation, the critical tunneling current is proportional to the barrier transparency $\lambda^{-1}$, and to satisfy the condition of a small $j$, low transparency is required, i.e., $\lambda \gg 1$.

In the case of an arbitrary current, the equation (36) with the barrier (38) can also be solved exactly [55, 56]. Instead of (39), (41) and (43), we now have the following expressions:

$$z(x) = 2j^2 + \left(1 - 2j^2\right)\tanh^2\left(\left(|x| + x_0\right)\sqrt{\frac{1 - 2j^2}{2}}\right), \qquad (45)$$

$$z_0^3 - \left(2 + 2j^2 + \frac{\lambda^2}{2}\right)z_0^2 + \left(1 + 4j^2\right)z_0 - 2j^2 = 0, \qquad (46)$$

$$\Delta\varphi = 2\left[\arctan\sqrt{\frac{1 - 2j^2}{2j^2}} - \arctan\sqrt{\frac{z_0 - 2j^2}{2j^2}}\right]. \qquad (47)$$

To obtain the relationship between the current $j$ and the phase difference $\Delta\varphi$, we need to solve the cubic equation (46). An analysis of this equation shows that one of its roots, which remains real for any $j$ and $\lambda \neq 0$, is always greater than one and must be discarded. Consequently, we must ensure that the given equation has more than one real root, which is equivalent to the requirement that the discriminant $D$ of equation (46) is non-negative, i.e.,

$$D = 2\left(1 - 2j^2\right)^3 \lambda^2 + \left(\frac{1}{4} - 10j^2 - 8j^4\right)\lambda^4 - j^2\lambda^6 \geq 0. \qquad (48)$$

The equation $D = 0$ determines the critical value of the current $j_{cr}$ as a function of $\lambda$ (see Fig. 4); due to its cumbersomeness, we do not present its analytical form here.

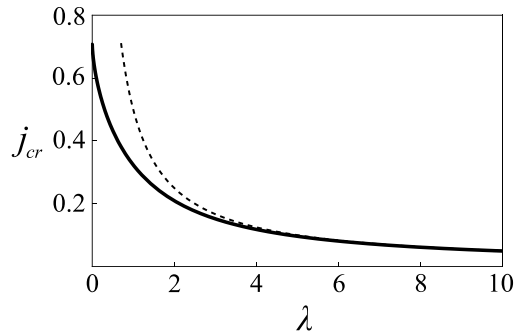

*Fig.4.* The dependence of the critical current $j_{cr}$ on $\lambda$. The solid line corresponds to the solution of equation (48), while the dashed line represents the relation $j_{cr} = 1/2\lambda$, which describes the limit of low transparency $\lambda \to \infty$.

In the limit of low (cf. (44)) and full transparency we have

$$j_{cr} = \frac{1}{2}\left(\frac{1}{\lambda} - \frac{1}{\lambda^3}\right), \quad \lambda \to \infty, \qquad (49)$$

$$j_{cr} = \frac{1}{\sqrt{2}}\left(1 - \frac{3}{4}\lambda^{\frac{2}{3}}\right), \quad \lambda \to 0. \qquad (50)$$

For the subcritical current $j \leq j_{cr}$ at a given value of $\lambda$, the remaining roots of equation (46) are also real and, according to (47), they determine the relationship between the current $j$ and the phase difference $\Delta\varphi$ (see Fig. 5).

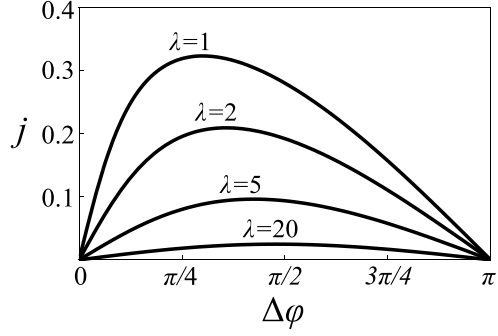

*Fig.5.* The dependence of the current $j$ on the phase difference $\Delta\varphi$ for different values of the parameter $\lambda$.

Returning to the case of small currents, we note that result (44) can be obtained in a simpler way by considering the fact that for $j \ll 1$, the dependence of the phase $\varphi$ on coordinates exhibits a step-like behavior (see Fig. 3(a)). Taking into account the nature of the function $f$ dependence on coordinates (39), the wave function $\Psi(x)$ can be approximated by the expression

$$\Psi(x) = \sqrt{n} e^{ijx} \tanh\left(\frac{|x|}{\sqrt{2}}\right) \exp\left(\operatorname{sgn}(x)\frac{i\Delta\varphi}{2}\right). \tag{51}$$

Taking into account the matching condition of the wave function at zero

$$\Psi(+0) = \Psi(-0) \equiv \Psi(0); \quad \frac{d\Psi(+0)}{dx} - \frac{d\Psi(-0)}{dx} = \lambda\Psi(0), \tag{52}$$

as well as the continuity of the current, we obtain the following expression:

$$j = \frac{i\hbar}{2j_0 M}\left[\Psi\frac{d\Psi^*}{dx} - \Psi^*\frac{d\Psi}{dx}\right] = \frac{i\hbar}{2j_0 M \lambda}\left[\frac{d\Psi(+0)}{dx}\frac{d\Psi^*(-0)}{dx} - \frac{d\Psi^*(+0)}{dx}\frac{d\Psi(-0)}{dx}\right]. \tag{53}$$

The substitution of the wave function (51) into (53) leads to expression (44).

An important feature of the obtained current (44), which corresponds to the low-density limit, compared to expression (25), which corresponds to the high-density limit, is that the critical current depends on the sum of the barrier potential amplitudes of the electron and hole layers, but not their product. This dependence follows from the Gross-Pitaevskii equation (32), where the layer potentials $U_\alpha(\mathbf{r})$ appear as a sum. This result implies that even in the absence of a barrier in one of the layers, the longitudinal Josephson effect will still occur due to the strong coupling of the pair components. If the barrier potentials in each layer are distant from each other, the phase of the order parameter will experience a jump at each barrier, thereby two successive Josephson junctions will take place. To quantitatively evaluate this effect, let us assume that the potential $U(x)$ in equation (36) is now represented as the sum of two delta-function potentials separated by a distance $l$ such that:

$$U(x) = U_{0e}\delta(x) + U_{0h}\delta(x-l). \tag{54}$$

Assuming that the distance $l$ significantly exceeds the coherence length $\xi$ and restricting ourselves to the case of a small current $j \ll 1$, the wave function $\Psi(x)$ can be approximated in a form analogous to (51)

$$\Psi(x) = \sqrt{n} e^{ijx}\left[\tanh\left(\frac{|x|}{\sqrt{2}}\right)\exp\left(\operatorname{sgn}(x)\frac{i\Delta\varphi_1}{2}\right)\theta\left(\frac{l}{2}-x\right) + \right.$$
$$\left. + \tanh\left(\frac{|x-l|}{\sqrt{2}}\right)\exp\left(\frac{i\Delta\varphi_1}{2} + \theta(x-l)i\Delta\varphi_2\right)\theta\left(x-\frac{l}{2}\right)\right], \tag{55}$$

where $\theta(x)$ is the Heaviside theta function. The quantities $\Delta\varphi_1$ and $\Delta\varphi_2$ represent the phase jump at each of the potential barriers so that the total phase jump equals $\Delta\varphi = \Delta\varphi_1 + \Delta\varphi_2$. Using the matching conditions of the type (52) for the wave function $\Psi(x)$ at points $x=0$ and $x=l$, as well as the definition of the current (53) and the condition of its continuity, we obtain

$$j = \frac{1}{2\lambda_1}\sin\Delta\varphi_1 = \frac{1}{2\lambda_2}\sin\Delta\varphi_2, \qquad (56)$$

where $\lambda_1 = U_{0e}/\gamma n\xi$, $\lambda_2 = U_{0h}/\gamma n\xi$. Taking into account that $\Delta\varphi = \Delta\varphi_1 + \Delta\varphi_2$, after straightforward algebraic transformations, we obtain from (56) the relationship between the tunneling current $j$ and the total phase difference $\Delta\varphi$

$$j = \frac{\sin\Delta\varphi}{2\sqrt{\lambda_1^2 + \lambda_2^2 + 2\lambda_1\lambda_2\cos\Delta\varphi}}. \qquad (57)$$

In particular, for the case $\lambda_1 = \lambda_2 = \lambda/2$, it follows from (57) that, in contrast to (44), the critical current will be equal to $1/\lambda$, corresponding to the critical phase difference $\Delta\varphi = \pi$.

### 4. Conclusions

The problem of non-dissipative longitudinal current states in bilayer systems with pairing of spatially separated electrons and holes in the presence of a potential barrier that divides the system into two macroscopic subsystems (right and left) was studied in the article. The flow of longitudinal currents through the barrier is naturally referred to as the longitudinal Josephson effect. The flow of current from the electron layer to the hole layer through an insulating interlayer is naturally referred to as the transverse Josephson current. Under experimental conditions, this current can be made very small, so in this work, the transverse Josephson current is neglected. It is shown that the dependence of the longitudinal current on the matrix elements of tunneling through the barrier is significantly determined by the degree of system dilution. In the case of high-density systems, where the size of the electron-hole pairs is much larger than the average distance between pairs, the current is proportional to the product of the matrix elements of tunneling through the barrier in the electron and hole layers. In the case of low-density systems, the current is proportional to the harmonic mean of these tunneling matrix elements. For high-density systems, the result was obtained by two methods: the tunneling Hamiltonian method and the *t*-representation method (see Appendix). In the general case, the coordinates of the potential barriers in the electron and hole layers may not coincide. When the coordinates of the barriers coincide, the current, for example, in the electron layer, is equal to $j_e = (1/2\lambda)\sin\varphi$, where $\lambda^{-1}$ is the total transparency of the barriers, and $\varphi$ is the phase drop of the order parameter at the barrier. The hole current $j_h = -j_e$. For low-density systems, in the case when the potential barriers in the electron and hole layers are shifted relative to each other by a distance that significantly exceeds the coherence length, the current is equal to $j_e = -j_h = \sin\varphi/2\sqrt{\lambda_e^2 + \lambda_h^2 + 2\lambda_e\lambda_h\cos\varphi}$. Here $\lambda_e^{-1}$, $\lambda_h^{-1}$ are the transparencies of the barriers in the electron and hole layers, and $\varphi = \varphi_e + \varphi_h$, where $\varphi_e$ and $\varphi_h$ are the phase drop at the barrier in the electron and hole layers, respectively. These results show that in systems with pairing of spatially separated electrons and holes, a non-dissipative electric current through the potential barrier separating the left and right sides of the system can flow.

# Appendix

The tunneling Hamiltonian method, used in Section 2, was proposed in 1962 in work [57]. In the same year, Josephson, using the tunneling Hamiltonian, predicted the possibility of a persistent superconducting current flowing through a tunnel contact formed by two superconductors. The results obtained within the framework of the tunneling Hamiltonian method agree well with experiments, but this method does not answer some fundamental questions. In particular, the model scheme does not allow including the tunnel current in the conventional formalism of current state descriptions in modern microscopic theory of superconductivity, where spatial phase variations of the order parameter play a significant role.

These questions stimulated the development of a consistent microscopic theory of tunneling current in superconductors. The starting point of this theory is Gorkov's equations. The next important step is to expand the Green's functions in terms of the eigenfunctions of the one-particle Schrödinger equation with a potential barrier, thus eliminating the barrier from Gorkov's equations. Finally, the equation for the order parameter and the expression for the current through the normal Green's function should be averaged over atomic-length scales. This procedure is known as the "*t*-representation method" [58].

A comparison of results obtained using the *t*-representation method and the model scheme (the tunneling Hamiltonian method) showed that the tunneling Hamiltonian is an effective Hamiltonian adapted for calculating the current at the barrier in the first order of perturbation theory with respect to the transparency.

In this Appendix, we apply the *t*-representation method to study the longitudinal tunnel effect in systems where superconductivity is due to the pairing of spatially separated carriers.

Let us write Gorkov's equations for the Fourier components of the normal $G$ and anomalous $F$ Green's functions

$$\begin{cases} \left( i\omega + \dfrac{1}{2m_e} \dfrac{\partial^2}{\partial x^2} + \mu_e - U_e(x) \right) G_\omega(x, x') + \Delta^*(x) F_\omega(x, x') = \delta(x - x') \\ \left( i\omega - \dfrac{1}{2m_h} \dfrac{\partial^2}{\partial x^2} - \mu_h + U_h(x) \right) F_\omega(x, x') + \Delta^*(x) G_\omega(x, x') = 0 \end{cases} \quad (A1)$$

Here, $\mu_e$ and $\mu_h$ are the chemical potentials of electrons and holes, respectively, and the system of units $\hbar = 1$ is used. In the equations (A1), the barrier potential in the electron $U_e(x)$ and hole $U_h(x)$ layers is introduced. To exclude these potentials, it is convenient to expand the functions $G_\omega$ and $F_\omega$ in terms of the eigenfunctions $\psi_p^{(i)}(x)$ and $\phi_p^{(i)}(x)$ (the index $i$ indicates from which side the free particle falls on the insulating layer) of the single-particle Hamiltonian with a barrier in the electron and hole layers, respectively (cf. [2])

$$\begin{cases} G_\omega(x, x') = \sum_{p,p',i,k} G_{p_1 p_2}^{ik} \psi_p^{(i)} \psi_{p'}^{*(k)} \\ F_\omega(x, x') = \sum_{p,p',i,k} F_{p_1 p_2}^{ik} \phi_p^{(i)} \psi_{p'}^{*(k)} \end{cases} \quad (A2)$$

In result, the equations (A1) take the following form:

$$\begin{cases} \left( i\omega - \xi_p^e \right) G_{pp'}^{ik} + \sum_{p_1, l} \left\langle \psi_p^{(i)} \left| \Delta^* \right| \phi_{p_1}^{(l)} \right\rangle F_{p_1 p'}^{lk} = \delta_{ik} \delta_{pp'} \\ \left( i\omega + \xi_p^h \right) F_{pp'}^{ik} + \sum_{p_1, l} \left\langle \phi_p^{(i)} \left| \Delta \right| \psi_{p_1}^{(l)} \right\rangle G_{p_1 p'}^{lk} = 0 \end{cases} \quad (A3)$$

The system of equations (A3) is valid in the most general case. However, we will consider a particular case when the dependence of the potential barrier on the coordinates is chosen in the form $U_\alpha(x) = U_{0\alpha}\delta(x)$, where $\delta(x)$ is the Dirac delta function, $\alpha = e, h$. In this case, the wave functions $\psi_p^{(i)}(x)$ take the form

$$\psi_p^{(1)}(x) = \frac{1}{\sqrt{2L}}\left\{\frac{p}{p+iK_1}e^{ipx} - \frac{2K_1\theta(-x)}{p+iK_1}\sin(px)\right\}, \quad (A4)$$

$$\psi_p^{(2)}(x) = \frac{1}{\sqrt{2L}}\left\{\frac{p}{p+iK_1}e^{-ipx} + \frac{2K_1\theta(x)}{p+iK_1}\sin(px)\right\}. \quad (A5)$$

Here, $2L$ is the longitudinal size of the system, $K_1 = m_e U_{0e}$, $p = \pi n/L$, $n \in \mathbb{Z}$. Expressions for the functions $\phi_p^{(i)}(x)$ are obtained from (A4) and (A5) by replacing $K_1$ with $K_2 = m_h U_{0h}$.

Assuming further that the order parameter $\Delta$ experiences a phase jump at the barrier

$$\Delta = |\Delta| \cdot \left(\theta(-x)e^{-i\varphi/2} + \theta(x)e^{i\varphi/2}\right), \quad (A6)$$

in the zeroth approximation in powers of transparency $1/m_\alpha U_{0\alpha}$ from (A3), (A4) and (A5) we obtain

$$G_{pp'}^{11} = G_{pp'}^{22} = \frac{i\omega + \xi_p^h}{(i\omega - \xi_p^e)(i\omega + \xi_p^h) - |\Delta|^2}\delta_{pp'}, \quad (A7)$$

$$F_{pp'}^{11} = -\frac{|\Delta|e^{-i\varphi/2}}{(i\omega - \xi_p^e)(i\omega + \xi_p^h) - |\Delta|^2}\delta_{pp'}. \quad (A8)$$

The expression for $F_{pp'}^{22}$ is given by formula (A8) with the substitution $\varphi \to -\varphi$. The off-diagonal elements in this approximation are equal to zero.

In the first order of transparency, from (A3), we obtain the following equations for the off-diagonal elements $G_{pp'}^{12}$ and $F_{pp'}^{12}$:

$$\begin{cases} (i\omega - \xi_p^e)G_{pp'}^{12} + \sum_{p_1}\langle\psi_p^{(1)}|\Delta^*|\phi_{p_1}^{(2)}\rangle F_{p_1 p'}^{22} + \sum_{p_1}\langle\psi_p^{(1)}|\Delta^*|\phi_{p_1}^{(1)}\rangle F_{p_1 p'}^{12} = 0 \\ (i\omega + \xi_p^h)F_{pp'}^{12} + \sum_{p_1}\langle\phi_p^{(1)}|\Delta|\psi_{p_1}^{(2)}\rangle G_{p_1 p'}^{22} + \sum_{p_1}\langle\phi_p^{(1)}|\Delta|\psi_{p_1}^{(1)}\rangle G_{p_1 p'}^{12} = 0 \end{cases}. \quad (A9)$$

The equations for $G_{pp'}^{21}$ and $F_{pp'}^{21}$ are obtained from (A9) by interchanging all upper indices ($1 \leftrightarrow 2$). The matrix elements involved in the system are computed directly using expressions (A4), (A5) and (A6). Substituting the Green's functions from (A7) and (A8) into the system of linear equations (A9), we obtain expressions for the off-diagonal elements of $G_{pp'}^{ik}$ and $F_{pp'}^{ik}$ ($i \neq k$). Since, for the calculation of the Josephson current, the expression for the $G$-function is required, we explicitly write the result only for $G_{pp'}^{ik}$:

$$G_{pp'}^{12} = -\frac{2pp'|\Delta|^2}{LK_1 K_2(p^2 - p'^2)}\frac{(K_2 - K_1\exp[i\varphi])(\xi_p^h - \xi_{p'}^h)}{(|\Delta|^2 - (i\omega - \xi_p^e)(i\omega + \xi_p^h))(|\Delta|^2 - (i\omega - \xi_{p'}^e)(i\omega + \xi_{p'}^h))}. \quad (A10)$$

The matrix elements $G_{pp'}^{21}$ and $G_{pp'}^{12}$ are equal up to the replacement $\varphi \to -\varphi$.

The current in the electron layer is calculated using the formula

$$I_e = \sum_{p,p',i,k} J_{p'p}^{ki} T\sum_\omega G_{pp'}^{ik}, \quad (A11)$$

where $J_{p'p}^{ki}$ is the matrix elements of the current operator by eigenfunctions $\psi_p^{(k)}$. Calculating the current at $x = 0$ up to the order $1/(m_\alpha U_{0\alpha})^2$ and using (A7), it is easy to see that the diagonal matrix elements of the current operator do not contribute to the total current $I_e$, since in this approximation they cancel each other out. The off-diagonal elements of the current operator at $x=0$ in the first non-zero approximation are equal to

$$J_{p'p}^{12} = -J_{p'p}^{21} = \frac{ie}{2m_e}\left(\psi_{p'}^{(1)}\frac{\partial \psi_p^{(2)*}}{\partial x} - \psi_p^{(2)*}\frac{\partial \psi_{p'}^{(1)}}{\partial x}\right) = \frac{ie}{2Lm_e K_1}pp', \quad (A12)$$

where $e > 0$ is the elementary electric charge. Substituting (A10), (A12) into (A11), and also using the fact that $|\Delta|^2 - (i\omega - \xi_p^e)(i\omega + \xi_p^h) = \varepsilon_p^2 + (\omega + i\eta_p)^2$, we obtain the following expression for the Josephson current in the *t*-representation method:

$$I_e = -\frac{e\sin\varphi|\Delta|^2}{L^2 m_e m_h K_1 K_2}\sum_{\mathbf{p},\mathbf{p'},\omega} T\frac{p^2 p'^2}{\left(\varepsilon_p^2 + (\omega + i\eta_p)^2\right)\left(\varepsilon_{p'}^2 + (\omega + i\eta_{p'})^2\right)}. \quad (A13)$$

The Josephson current obtained by the tunneling Hamiltonian method is given by equation (25). Expressions (A13) and (25) coincide if

$$T_{p,p'}^\alpha = \frac{pp'}{\sqrt{2}Lm_\alpha^2 U_{0\alpha}}. \quad (A14)$$